\documentclass[12pt]{article}
\usepackage{url}
\usepackage[hidelinks]{hyperref}
\usepackage{amsmath}
\usepackage{multirow}
\usepackage{hhline}
\usepackage{graphicx}
\usepackage{float}
\usepackage{caption}
\usepackage{subfigure}
\usepackage{textcomp}
\usepackage{dblfloatfix}

\begin{document}

\section{Introduction}\label{introduction}

Web attacks are attacks that exclusively use the HTTP/HTTPS protocol.
Symantec Internet Security Threat Report (ISTR) \cite{ISTR2016} reveals that the number of web attacks explosively increased, reaching 1.1 million every day in 2015, more than twice the rate in 2014 (0.493 million).
Among web attacks, code injection attacks against web applications \cite{Fonseca2014Evaluation} increased each year and accounted for at least 96.15\% of web attacks in 2015, according to the Imperva Web Application Attack Report (WAAR) \cite{WAAR2014,WAAR2015,WAAR2016}.
In 2015, Team GhostShell claimed to have hacked numerous websites using SQL injection (SQLI) attacks \cite{lawal2016systematic}, and disclosed thousands of compromised account details, including emails, user names, addresses, telephone numbers, and other privacy information \cite{Ghost}.

This paper focuses on the most frequent types of web-based code-injection attacks in 2015 \cite{WAAR2016}, namely, Cross-Site Scripting (XSS) attack \cite{hydara2015current}, SQLI attack, Directory Traversal (DT) \cite{WAAR2016}, and Remote File Inclusion (RFI) \cite{WAAR2016}, respectively accounting for 49.09\%, 28.32\%, 9.82\% and 8.92\% of the entire web attacks.
Attack vectors of these attacks exist in user input.
Whenever user input is improperly handled, these attacks can succeed.
Since most user input data exists in queries \footnote{Web request query string is consistently termed as query in this paper for simplicity.}, detecting malicious queries is the core of detecting web-based code injection attacks.
For web applications whose source code is unavailable or difficult to obtain, Web Intrusion Detection Systems (Web IDSs) are the only option \cite{Prokhorenko2016Web}.
A web IDS acts as an intermediate layer between protected web applications and users, and analyzes web traffic to detect possibly malicious activities.

A considerable amount of effort has been made to detect malicious queries in web requests using web IDSs.
Two major detection approaches are utilized: signature-based detection and anomaly-based detection.
A popular signature-based detection approach is WAF, which maintains a signature library of known signatures, and compares new web requests with the signatures.
Signature-based approaches are efficient in detecting known attacks with a low False Positive (FP) rate, but unable to detect previously unknown attacks(e.g., zero-day attacks).
Anomaly-based detection approaches \cite{Kruegel2003Anomaly,Kruegel2005A,Robertson2006Using,Song2009Spectrogram,Kozakevicius2015URL,Juvonen2015Online} usually rely on a detection model to identify anomalous web requests.
In contrast to signature-based approaches, anomaly-based approaches can detect unknown attacks, but with a high FP rate.
The two approaches are often applied in a complementary fashion in web IDSs: anomaly detection serves to discover unknown attacks, whose attack signatures are then utilized by signature-based approach for detection.
The ability to detect unknown attacks has drawn wide academic attention to anomaly-based approaches.

However, existing anomaly-based approaches for web attack detection are constant, in other words, they usually collect all training data once to induce a constant detection model.
Since attackers have become more sophisticated and utilized more advanced web attack toolkits, the detection model might become obsolete and outdated, incapable of detecting the latest malicious queries in web attacks.
Previous adaptive attack detection methods \cite{Meng2014Adaptive,Wang2014Autonomic,Zhang2015Detecting} are designed for network intrusions and are not applicable to web attacks.
A practical solution for keeping the web attack detection model constantly updated is to incorporate the latest important queries, including informative benign queries and representative malicious queries.
The existing approach to obtaining the latest important queries is randomly selecting requests from web traffic and then manually labeling them.
Unfortunately, the majority of randomly selected queries are similar benign ones, and there is low probability of a query being important.
Since web traffic is huge, it does take considerable manual labeling work to obtain important queries.
This challenge has motivated our research work to build a malicious query detection system that can adaptively incorporate the latest important queries to update the detection model.

In this paper, we address the issues related to the adaptive detection of malicious queries in web attacks.
We present AMODS, an adaptive system for this purpose.
AMODS leverages an efficient adaptive learning strategy, SVM HYBRID, which is a hybrid of Suspicion Selection (SS) and Exemplar Selection (ES).
The former features in acquiring the most important informative queries, namely, suspicions, while the latter specializes in harvesting representative malicious queries, exemplars.
Suspicions and exemplars are called important queries.
AMODS aims to identify attacks as early as possible by periodically checking the latest important queries.
The number of important queries is trivial, so manual labeling work is reduced to the minimum.
The important queries are then incorporated into the training pool to update the detection model, which is a stacking-based ensemble classifier, composed of three base classifiers and a meta classifier.
SVM \cite{Vapnik2006Estimation} is used as its meta classifier, so that SVM HYBRID can exert the detection model to obtain important queries.
A ten-day query dataset collected from an academic institute website's web server logs is used for evaluation.
Since the logs record requests passing through a commercial WAF, malicious queries obtained by our system can be used to update the WAF signature library.

The main contributions of this study are as follows:

\begin{itemize}
\item We present AMODS, an adaptive system for detecting malicious queries in web attacks. To the best of our knowledge, this is the first work in the adaptive detection of web attacks.

\item We propose SVM HYBRID, an adaptive learning strategy designed for the efficient selection of important queries.

\item We employ stacking, a meta-learning \cite{brazdil2008metalearning} based ensemble classifier, as the underlying detection model for accurate malicious query detection.

\item AMODS outperforms existing web attack detection methods with an F-value of 94.79\% and FP-rate of 0.09\% on a real-world ten-day query set. The total number of malicious queries obtained by SVM HYBRID is 2.78 times that by SVM AL during the ten-day detection using AMODS.
\end{itemize}

The remainder of the paper is organized as follows.
Section \ref{Section2.backRela} provides background and related work.
Detailed design and implementation of the system, together with the proposed adaptive learning strategy, namely, SVM HYBRID, will be presented in Section \ref{Section3.design}.
Section 4 shows data collection and preprocessing.
Next, Section \ref{Section4.experiment} introduces the process of the determination of feature reduction and stacked classifiers, then shows the efficacy of our system.
In addition, AMODS is compared with constant models and adaptive models, including SVM AL and random selection, in terms of detection performance and the number of malicious queries obtained.
Detection performance comparisons with related work are also demonstrated.
Afterwards, we clarify some relevant issues of our approach in Section \ref{Section5.discussion}.
Finally, Section \ref{Section6.conclusion} concludes this paper.

\bibliographystyle{plain}

\bibliography{mybibfile}

\end{document}


\section{Background and related work}\label{Section2.backRela}

\begin{figure}[b]
    \centering
    \includegraphics{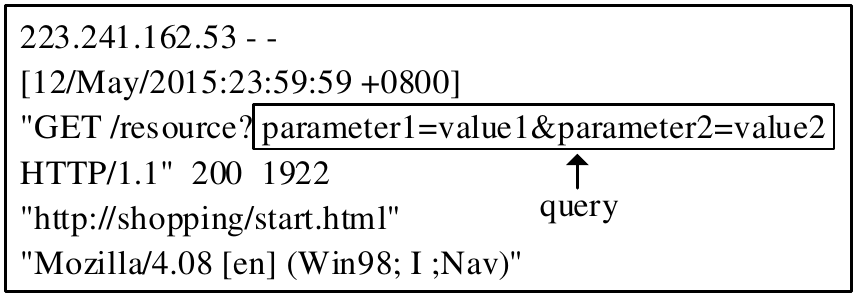}
    \caption{Web server access log entry}
    \label{fig:log}
\end{figure}

\begin{table*}
\centering
\caption{Comparison of normal queries and malicious queries in web-based code injection attacks}
\label{tab:t1.querieComparison}
\begin{tabular}{ccl} \hline
No. &  Type   &  Query in a web request URL \\
\hline
(a) &  Normal &  http://www.victim.com/index.php?postID=123 \\
\multirow{2}{*}{(b)} & \multirow{2}{*}{ SQLI}   &  http://www.victim.com/index.php?postID='/**/union/**/select/**/0, \\
  & & concat(username,0x3a,password)/**/from/**/users/*' \\
\multirow{2}{*}{(c)} &  \multirow{2}{*}{ XSS}    &  http://www.victim.com/index.php?postID=$<$script$>$document.location \\
 & & =``http://vicious\_site/stealcookie.cgi?''+document.cookie$<$/script$>$ \\
(d) &  DT     &  http://www.victim.com/index.php?postID=../../../../etc/passwd \\
(e) &  RFI    &  http://www.victim.com/index.php?postID=http://vicious\_site/hack.txt?ls \\
\hline
\end{tabular}
\end{table*}

\begin{figure*}[b]
    \centering
    \includegraphics[width=1.7\columnwidth]{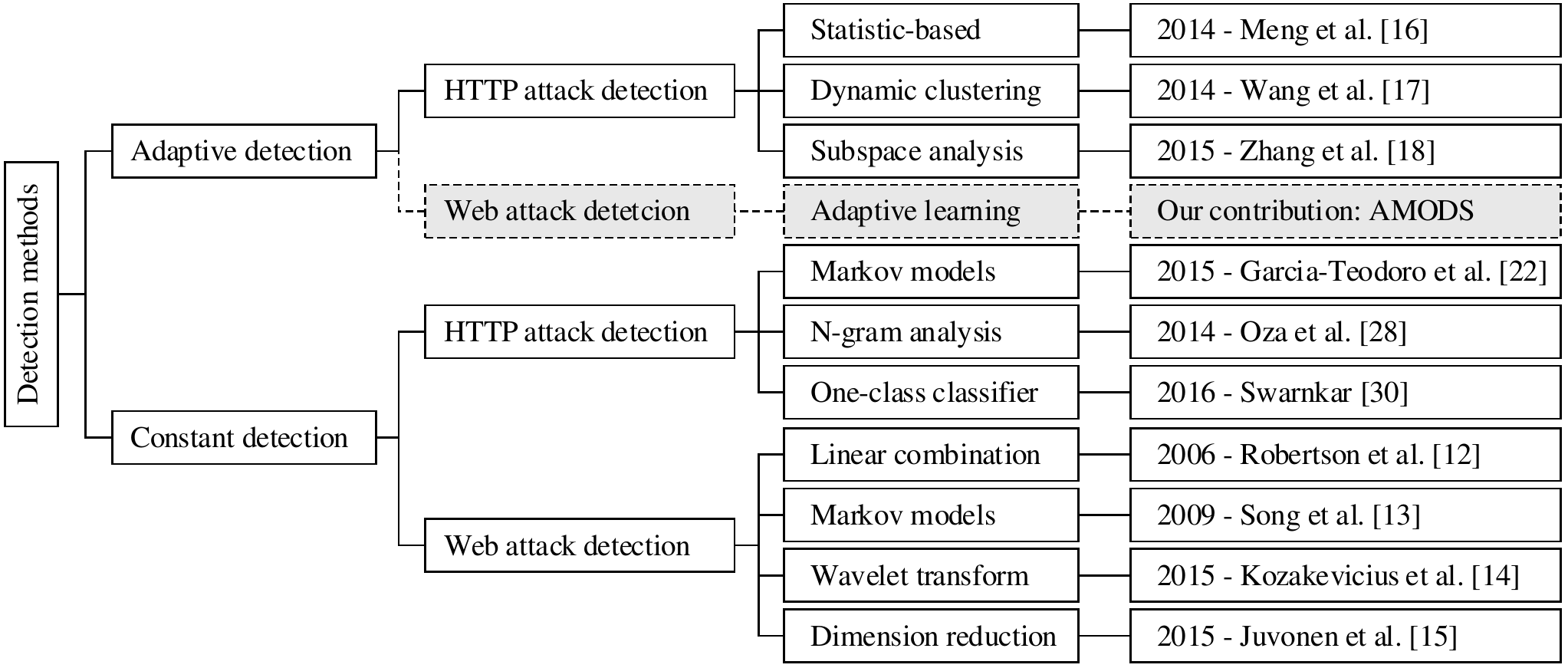}
    \caption{Taxonomy of the latest academic research on anomaly detection methods of web attacks and HTTP attacks}
    \label{fig:taxonomy}
\end{figure*}

\subsection{Background}

Queries contain most user inputs and are thus closely related to web-based code injection attacks, which are the most threatening web-layer attacks \cite{WAAR2014,WAAR2015,WAAR2016}.
Therefore, our work concentrates on malicious query detection in order to detect web-based code injection attacks.
Queries can be found in HTTP GET requests recorded in web server logs.
Figure \ref{fig:log} shows a web server log entry in Common Log Format (CLF), consisting of various items, such as the source IP address, timestamp, sever response status code, user agent header field, and the actual web request string.
The framed text in the request string is a query, which our system examines.
A query is identified by a leading ``?" character following the referenced resource, and lists pairs of parameter names and values.

Malicious queries in web-based code injection attacks and normal queries are different in character distribution and character sequences.
Table \ref{tab:t1.querieComparison} compares a normal request and four types of code injection attacks analyzed in this paper: SQLI attack, XSS attack, DT attack, and RFI attack.
We include Remote Code Execution (RCE) into XSS attack, since RCE is essentially a special kind of XSS attack \cite{Zheng2013Path}.
The requests in Table \ref{tab:t1.querieComparison} are forwarded to a vulnerable script named ``index.php" in a forum website, and the script can extract posts from the database and present a post to a user.
``postID" is a parameter name for this script and takes an integer input value as the post ID while recording the transaction.
As shown in Table \ref{tab:t1.querieComparison}, (a) ``postID=123" is a normal query, and (b)-(e) are malicious queries.
(b) is a typical SQLI attack that attempts to obtain confidential data from a restricted table.
(c) conducts an XSS attack to obtain user cookies using a Common Gateway Interface (CGI) script on the attacker host.
(d) conducts a DT exploit to discover the file ``passwd" using string ``../../../".
Since ``passwd" is a local file on the vulnerable server, (d) is also called Local File Inclusion (LFI) attack.
(e) tries to load ``hack.txt" file, a bash shell, from a remote malicious sever, and execute the ``ls" command within the bashshell in an RFI attack.

\subsection{Related work}

Applying anomaly detection to attack detection has been an active research area.
Based on whether a detection method can adapt to the change of web attacks, we divide existing web attack and HTTP attack detection techniques that use anomaly detection schemes into two categories: adaptive detection and constant detection.
Adaptive detection methods always incorporate new traffic data to induce an up-to-date detection model, while constant detection methods use a period of traffic data to build a detection model once and use it for all later detection.

Unfortunately, existing adaptive detection techniques \cite{Meng2014Adaptive,Wang2014Autonomic,Zhang2015Detecting} are designed for network intrusion detection.
They might not be suitable for web attack detection, since the landscape of network intrusions differs from web attacks.
Malicious queries in web attacks have more subtle distinctions from legitimate queries in normal requests.
Constant detection methods \cite{Kruegel2003Anomaly,Kruegel2005A,Robertson2006Using,Song2009Spectrogram,Kozakevicius2015URL} cannot detect the latest unknown attacks.
The adaptive detection of web attacks has been found deficient.

In this paper, AMODS is proposed to bridge the gap between adaptive detection and web attack detection.
As far as we know, this is the first research in the adaptive detection of web attacks.
Figure \ref{fig:taxonomy} gives our taxonomy of the latest research on methods that use anomaly detection to detect web attacks as well as HTTP attacks.

\subsubsection{Adaptive detection}

Adaptive detection can detect the latest unknown attacks by adapting to the changing behaviors of attacks, yet adaptive detection methods to date have been only related to network intrusion detection.

\emph{HTTP attack detection.}
These detection methods take network traffic of port 80, namely HTTP traffic, as data source and analyze HTTP packet payloads.
Three types of unsupervised learning techniques are utilized:
(a) Statistic based.
Meng et al. \cite{Meng2014Adaptive} used a statistic-based approach to develop an adaptive blacklist-based packet filter, which can help filter out network packets based on IP confidence.
(b) Dynamic clustering.
Wang et al. \cite{Wang2014Autonomic} employed Affinity Propagation (AP) in dynamical clustering to detect HTTP attacks adaptively.
(c) Subspace analysis.
Zhang et al. \cite{Zhang2015Detecting} proposed Adaptive Stream Projected Outlier deTector (A-SPOT), which used a novel adaptive subspace analysis approach to detect anomalies from network connections.

Our proposed system AMODS falls into the category of adaptive methods for web attack detection.
AMODS leverages an efficient adaptive learning strategy, called SVM HYBRID.
Queries obtained by SVM HYBRID are used to update the detection model periodically.
AMODS can identify attacks at an early stage with a high F-value, and significantly improve the detection performance of the model over time.

\subsubsection{Constant detection}

Constant detection methods are widely applied to intrusion detection.
Though effective, they suffer from a high FP rate.
To deal with this problem, we combined SVM HYBRID with a stacking-based detection model.
Using our detection model, we achieved the lowest FP rate in comparison with above constant detection methods as shown in Section \ref{Section4.4.Experim4}.
Constant detection methods for HTTP attacks and web attacks are summarized as follows.

\emph{HTTP attack detection.}
Constant detection methods of HTTP attacks model packet payload in HTTP traffic, and operate at packet layer.
Four categories of detection algorithms are used:
(a) Markov models.
Garcia-Teodoro et al. \cite{Garcia2015Automatic} represented the normal behaviors of a service by Markov models, aiming to generate HTTP intrusion signatures for Network Intrusion Detection Systems (NIDS).
Soon, Zhong et al. \cite{Zhong2015Detecting} used Markov models whose inputs are each character of parameter values in normal payloads to build the normal profile.
Earlier works, such as HMMPayl \cite{Ariu2011HMMPayl}, made similar use of Markov models.
(b) N-gram analysis.
N-gram is a text categorization technique and is widely used in NIDSs, such as PAYL \cite{Wang2004Anomalous}, Anagram \cite{Wang2006Anagram}, RePIDS \cite{Jamdagni2013Review} and \cite{Oza2014HTTP}.
(c) One-class classifier.
McPAD \cite{Perdisci2009McPAD} detects HTTP attacks using multiple one-class SVMs, whose outputs are combined to make a final prediction.
Recently, Swarnkar \cite{Swarnkar2016OCPAD} adopted a one-class Multinomial Naive Bayes classifier and used likelihood of each short sequence¡¯s occurrence in payloads of known benign packets as a measure to derive the degree of maliciousness of a packet.

\emph{Web attack detection.}
Constant detection methods of web-layer code injection attacks usually analyze queries in web traffic (web server logs), and operate at application layer.
Two detection methods are mainly used:
(a) Linear combination.
Linear combination method creates a profile for normal traffic and considers new web requests with certain noticeable deviations from the normal profile as attacks.
Kruegel and Vigna \cite{Kruegel2003Anomaly,Kruegel2005A} were the first to analyze web server logs to detect attacks against web servers and web applications.
They calculated the final anomaly score for each web request using the linear combination of six models, such as the character distribution of query parameters.
Robertson et al. \cite{Robertson2006Using} used similar models to generalize suspicious web requests into anomaly signatures.
(b) Markov models.
Spectrogram \cite{Song2009Spectrogram} used a linear mixture of multiple Markov-chains to model queries.
(c) Wavelet transform.
Kozakevicius et al. \cite{Kozakevicius2015URL} proposed a URL query string anomaly sensor based on the properties of the bidimensional Haar wavelet transform.
(d) Dimensionality reduction.
Juvonen et al. \cite{Juvonen2015Online} proposed an anomaly detection system to detect anomalies in queries, and used dimensionality methods for data preprocessing.

It is worth mentioning that we only concentrate on adaptive detection of web attacks; two related problems, namely vulnerability exploitation and signature generation, are beyond the scope of this paper.
On the one hand, the criterion of query labeling is a combination of signature-based method and manual analysis, as described in Section \ref{Section5.discussion}, regardless of whether the corresponding requests successfully exploit the vulnerabilities of an application or not.
Web application vulnerability exploitation detection can be done via either static analysis \cite{Dahse2014Static} or dynamic analysis, such as taint tracking \cite{Livshits2010Towards} and taint inference \cite{Naderi2015Joza}.
On the other hand, we simply expect to obtain as many malicious queries as possible to update the WAF signature library.
Signature generation is outside our subject, which commonly uses regular expressions \cite{Howaed2014psigene}.

\bibliographystyle{plain}

\bibliography{mybibfile}


\begin{figure*}[t]
    \centering
    \includegraphics[width=1\columnwidth]{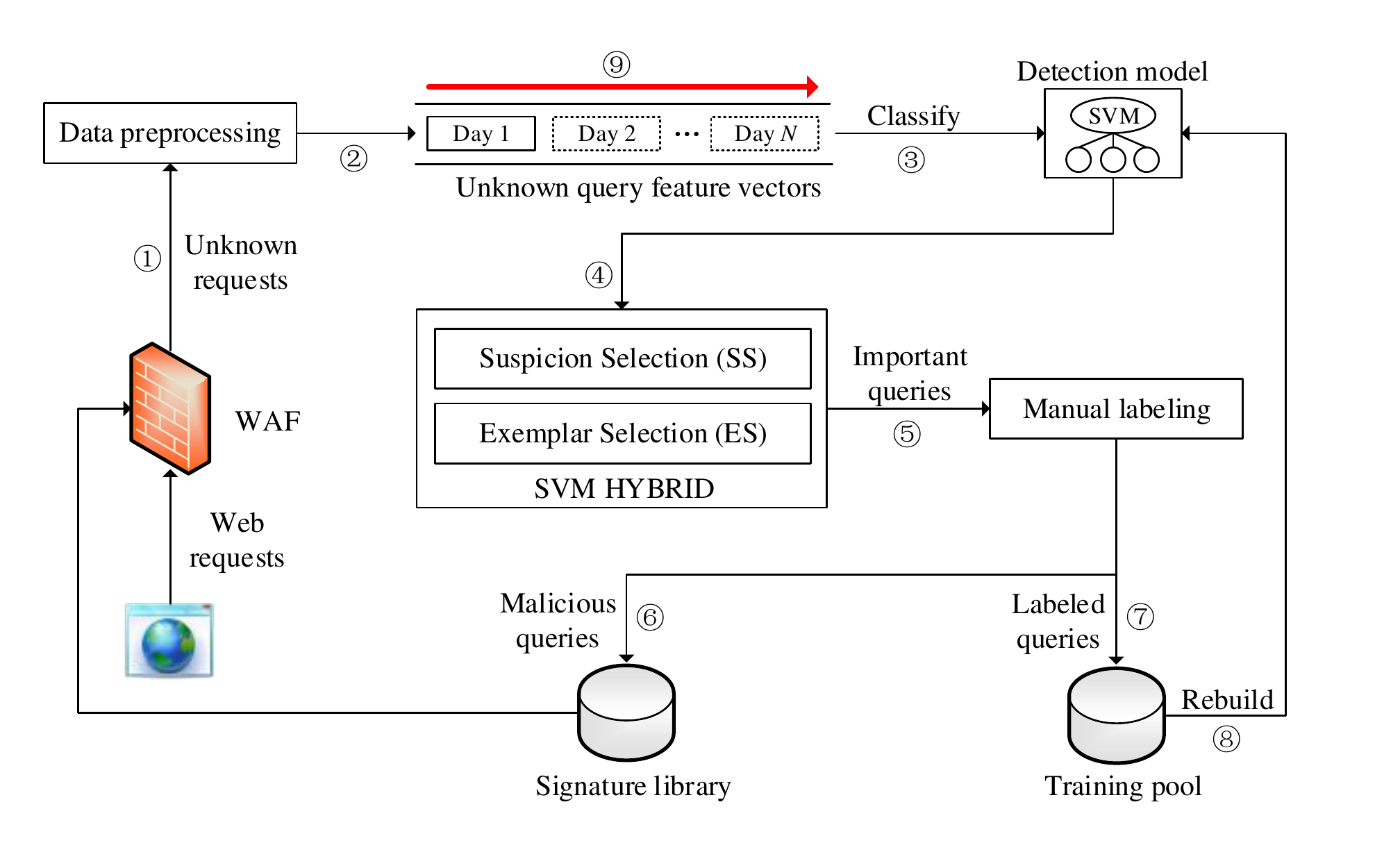}
    \caption{System overview}
    \label{fig:system}
\end{figure*}

\section{Design and implementation}\label{Section3.design}

In this section, we first illustrate the design of our Adaptive Malicious Query Detection System (AMODS, ``O" for ``Q"), followed by the core adaptive learning technique, named SVM HYBRID.
Lastly, we introduce our stacking-based training and classification.

\subsection{System design}\label{section3.1.System}

In this paper, we introduce the concept of \emph{important queries}, which consist of the most informative normal queries and the most representative malicious queries, as mentioned in Section \ref{introduction}. The goal of AMODS is to identify attacks as early as possible by periodically checking the latest important queries.
To reach this goal, our system leverages an adaptive learning approach called SVM HYBRID to obtain important queries.

The system is initialized with a training pool composed of a small set of labeled queries, from which the initial detection model is trained.
The detection model is based on stacking, and composed of three base classifiers and a meta classifier.
SVM is used as its meta classifier, so that SVM HYBRID could exert the detection model to obtain important queries.
To update the initial detection model, the system takes web server logs as its data source, and examines queries in HTTP GET requests in batch mode.
These queries are called unknown queries, since their class labels are unknown.
Classified unknown queries are analyzed by SVM HYBRID to obtain the smallest number of important queries for labeling.
Thus, the manual work is greatly reduced.
As shown in Figure \ref{fig:system}, the system proceeds as follows.

\begin{enumerate}
\item Collecting unknown queries.
Unknown queries are collected from $N$-day web requests.
Since attacks in the web requests have been filtered out by WAFs, the collected unknown queries contain benign queries and malicious queries undetected by WAFs.
\item Data preprocessing (Section \ref{Section3.3.Datapreparation}).
Data preprocessing transforms raw queries into feature vectors in batches of $N$ days, including data preparation, feature construction, and feature reduction.
\item Classification (Section \ref{Section3.4.training}).
Unknown queries on the $1^{st}$ day are classified using the current detection model.
\item Queries selection (Section \ref{Section3.2.SVMHYBRID}).
A fixed number of important queries are obtained from classified unknown queries on the $1^{st}$ day using SVM HYBRID.
\item Label the obtained important queries manually.
\item Update the WAF signature library using the obtained important queries that are labeled as malicious.
\item Add all the labeled important queries into the training pool\footnote{In following text, the terms ``training pool", ``training set", and ``labeled set" are used interchangeably.} .
\item Update the detection model (Section \ref{Section3.4.training}).
\item Iterate steps (3)-(8) for unknown queries in the rest $N-1$ days.
\end{enumerate}

\subsection{SVM HYBRID}\label{Section3.2.SVMHYBRID}

SVM HYBRID, our proposed adaptive learning strategy, obtains important queries to update the malicious query detection model.
SVM HYBRID operates in the kernel feature space of SVM (Support Vector Machine), and relies on positions of query samples in SVM kernel feature space to decide queries for labeling.
SVM has performed extremely well in many domains, particularly those involving text classification.

Given a domain $X$, a linear classifier used to separate positive and negative samples \cite{Vapnik2006Estimation} is defined in terms of a hyperplane
\begin{equation}\label{e1}
w^{T}x+b=0,\left( w\in R^{N},b\in R\right),
\end{equation}

\noindent where $w$ is the orthogonal normal vector to the hyperplane and is assumed to point towards the positive class, and $b$ is the bias of the hyperplane.
The linear classifier is corresponding to the decision rule

\begin{equation}\label{e2}
y=sgn(w^{T}x+b).
\end{equation}

Given training set $T=\{\left(x_1,y_1\right),\left(x_2,y_2\right),\dots,\left(x_n,y_n\right)\}$, where $x_i\in R^N$ is a feature vector, and $y_i\in \{-1,1\}$ is a class label, the margin is defined as the maximal perpendicular distance between samples of the two classes.
SVM aims to find the optimal linear classifier whose separating hyperplane maximizes the margin.
$w$ of the optimal linear classifier is solved as
\begin{equation}\label{e3}
w = \sum _{i=1}^n a_i y_i x_i .
\end{equation}

When original input samples are non-linearly separable, the samples could be mapped to a new kernel feature space where they are linearly separable.
Given a feature mapping process $\varphi$, the corresponding kernel is defined as
\begin{equation}\label{e4}
K(x,z) =\varphi \left( x\right) ^{T}\varphi \left( z\right).
\end{equation}

Using Equation (\ref{e3}) and (\ref{e4}), the decision value in Equation (\ref{e2}) can be rewritten as:
\begin{equation}\label{e5}
f(x) =\sum _{i=1}^{n}\alpha _{i}y_{i}K\left( x_{i},x\right) + b.
\end{equation}

\noindent $f(x)$ is the kernel distance between a sample $x$ and the hyperplane.
Calculating $K(x,z)$ is usually trivial.
Besides, according to the Karush-Kuhn-Tucker (KKT) conditions, $\alpha_i$ is non-zero only for samples that are closet to the hyperplane, namely support vectors, the number of which is usually small.
Therefore, the kernel distance is efficient to compute.

For data that are non-linearly separable though mapped to infinite dimensional feature space, the optimization problem can be formulated as
\begin{align}\label{e6}
\begin{split}
max _{\alpha } \quad  & \sum _{i=1}^{n}\alpha _{i}-\dfrac {1} {2}\sum _{i,j=1}^{n}\alpha _{i}\alpha _{j} y_i y_j x_i^T x_j\\
s.t.  \quad  & 0 \leq \alpha _i\leq C,i=1,\dots,n\\
& \sum _{i=1}^{n}\alpha _{i} y_i = 0,
\end{split}
\end{align}

\noindent where penalty parameter $C$ controls the relative weighting of a large margin and low misclassification rate of training data.
A small $C$ prefers a large margin, at the cost of a large number of misclassified training samples, corresponding to a soft margin classifier.
By contrast, a large $C$ means a hard margin classifier.

\begin{figure}
    \centering
    \includegraphics[width=0.7\columnwidth]{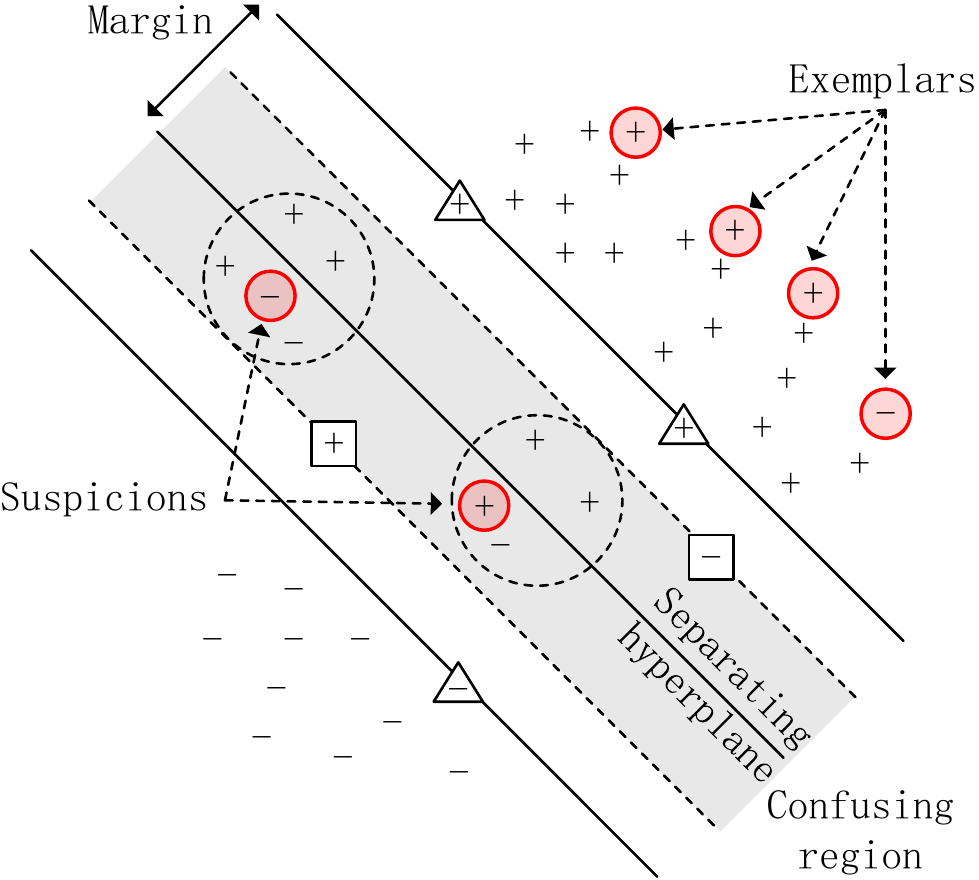}
    \caption{Illustrative graph of SVM HYBRID}
    \label{fig:SVMHYBRID}
\end{figure}

The illustrative graph of SVM HYBRID is presented in Figure \ref{fig:SVMHYBRID}.
The middle solid line is the separating hyperplane, while the two external solid lines indicate the margin.
The upper side of the hyperplane is the malicious side.
``$+$" represents a query in the positive class, namely, a malicious query, while ``$-$" stands for a query in the negative class: a normal query.
SVM HYBRID comprises of Suspicion Selection (SS) and Exemplar Selection (ES).
\emph{SS obtains the most important informative queries}, which are suspected of being malicious and thus are called suspicions.
Suspicions are within the prone-to-be-misclassified region, called the confusing region: the grey area delimited by two dashed lines in Figure \ref{fig:SVMHYBRID}.
The two points surrounded by red circles within the confusing region are suspicions.
Conversely, \emph{ES obtains the most representative malicious queries}, which are called exemplars, since each of them is an exemplar of similar malicious queries.
Exemplars lie in the malicious side and far away from the hyperplane.
As shown in Figure \ref{fig:SVMHYBRID}, the four points surrounded by red circles on the upper side are exemplars.
The main contribution of suspicions and exemplars is also different: the former contributes more to detection performance improvement, while the latter is conducive more to harvesting truly malicious queries.
The contribution of SVM HYBRID is a hybrid of these strengths.

SS and ES operate on two disjointed subsets of each unknown query set. SVM HYBRID starts with SS on one subset and then switches to ES on the other subset.
The ratio of the size of query subset for SS to that for ES is denoted as $\theta$.
$\theta$ is an adjustable parameter in SVM HYBRID to address the tradeoff between detection performance improvement and the number of malicious queries obtained.
The effects of $\theta$ set to different values are evaluated in Experiment 2 (Section \ref{Section4.2.Experim1}).

Moreover, SVM HYBRID is suitable for SVMs with any kernels.
Radial basis function (RBF) kernel and Polynomial kernel, two widely used kernels, are tested in our experiment, respectively defined as
\begin{align}\label{e7}
K(x,y) & = e ^{-\gamma\|x-y\|^2}  \\
K(x,y) & = \left( \left(x \cdot y \right) + \theta \right)^d.
\end{align}

\subsubsection{Suspicion Selection (SS)}\label{Section3.2.1.SS}

Adaptive learning with SVM uncertainty sampling is also known as SVM active learning (we call it SVM AL in the rest of this paper) \cite{Schohn2000Less,Tong2001Support}.
SS is similar to SVM AL, since they all select unlabeled samples within the margin.
However, SVM AL simply selects samples closet to the hyperplane and does not consider their distribution.
Consequently, similar samples might be selected, making SVM AL prone to sample bias.
Labeling similar samples is also a waste of manual work.
Therefore, given a large pool of unlabeled samples, it is desirable to preserve the density distribution of samples in the pool during the learning process.

Our proposed Suspicion Selection (SS) pursues the idea of preserving data distribution by using clustering to find potential informative queries from unknown queries within the confusing region.
In brief, SVM AL concentrates on the most informative data, while SS focuses on the most important informative data so that data redundancy is reduced to the minimum.
The procedure of SS is as follows.

Firstly, the confusing region is determined.
The set of misclassified queries in a training set that fall into the margin during training is denoted as $Q$.
Queries in $Q$ are sorted in order according to their kernel distances computed using Equation (\ref{e5}).
The query in $Q$ with the minimal kernel distance (the ``$+$" point with the rectangular frame in Figure \ref{fig:SVMHYBRID}) determines the lower boundary of the confusing region (the lower dashed line).
Oppositely, the query in $Q$ with the maximal kernel distance (the ``$-$" point with the rectangular frame) determines the upper boundary of the confusing region (the upper dashed line). That is,
\begin{align}\label{e8}
f_{lower} & =\min _{x\in Q}f(x) \\
f_{upper} & =\max _{x\in Q}f(x).
\end{align}

The confusing region is defined as
\begin{equation}\label{e9}
{Confusing\thickspace region} \Leftarrow f_{lower} \leq f(x) \leq f_{upper}.
\end{equation}

To ensure the existence of the confusing region, the margin should be wide enough to allow misclassifications of training samples within the margin.
That is, the meta classifier SVM of our detection model should be a soft margin classifier.
The training misclassification rate and detection performance of the detection model can be balanced by adjusting the penalty parameter $C$ and SVM kernel parameter of the meta classifier SVM.

Secondly, $K$-medoids clustering \cite{Duda1973Pattern} is performed on the unknown queries that fall into the confusing region.
The final cluster centers are suspicions.
$K$-medoids begins with $K$ randomly selected samples as medoids (cluster centers).
Then each remaining sample is assigned to the cluster whose medoid is the closet to it.
That is, the partitioning method is based on the principle of minimizing the sum of the distances between each sample and its corresponding referenced medoid.
This sum is denoted as $E$, and is defined as
\begin{equation}\label{e10}
E = \sum _{j=1}^k \sum _{x \in C_j} |\varphi(x) - \varphi(M_j)|,
\end{equation}

\noindent where $x$ is a given sample in cluster $C_j$, whose cluster center is $M_j$.
$|\varphi (x)-\varphi(M_j)|$ is the Euclidean distance between $x$ and $M_j$ in kernel feature space, and using Equation (\ref{e4}), it can be given by
\begin{multline}\label{e11}
|\varphi(x) - \varphi(M_j)| \\ =\sqrt{K(x,x) + K(M_j,M_j) - 2K(x,M_j)}.
\end{multline}

The $K$ cluster centers with which $E$ reaches its minimum value are optimal.
The partitioning step is repeated until the number of unknown samples within the confusing region is less than $K$.
When partitioning terminates, the final cluster centers are the most important informative samples, called suspicions.
Suspicions are important samples whose class labels cause the most confusion to the classifier.

Figure \ref{fig:SVMHYBRID} shows that the confusing region covers two clusters, each represented by a dashed circle.
Their cluster centers are the two points surrounded by red circles in the two dashed circles.
One cluster center is a normal query and another is a misclassified malicious query.
In other words, suspicions might be misclassified and also might not be malicious.
On the contrary, queries obtained by Exemplar Selection (ES) are highly likely to be malicious.
ES is described in the following subsection.

\subsubsection{Exemplar Selection (ES)}\label{Section3.2.2.ES}

Exemplar Selection (ES) serves to obtain representative malicious queries, called exemplars, from the unknown queries.
New web attacks are very commonly variants of existing attacks.
Queries in variant attacks might be similar to queries in existing attacks.
Labeling similar queries would unnecessarily increase manual labeling workload.
As a result, we hope that each exemplar is the unknown malicious query that differs most from all the malicious queries in the current labeled set.

\begin{table*}[t!]
\centering
\caption{Description of web server logs and data preprocessing}
\label{tab:t2.dataPreprocess}
\begin{tabular}{lccccccc} \hline
\multirow{3}{*}{Item}   & \multicolumn{2}{c}{Raw logs}  & \hspace{0.1cm} &\multicolumn{4}{c}{Data preprocessing}\\
\hhline{~--~----}    & Duration & Log size &\hspace{0.1cm} & Original & Cleaned & Normalized & Filtered \\
\hhline{~~~~~~~~} & (days) & (GB) &\hspace{0.1cm} & requests & queries & queries & queries \\
\hline

\# & 10 & 6.11 & \hspace{0.1cm} & 476,459 & 188,290 & 118,928 & 112,397 \\

\hline
\end{tabular}
\end{table*}

\begin{table}[t]
\centering
\caption{Statistics of query dataset}
\label{tab:t3.dataset_statistics}
\begin{tabular}{ccc} \hline
 \#Queries & Initial set & Ten unknown sets\\

\hline
Benign & 80 & 99,080 \\
Malicious & 20 & 920 \\
Total & 100 & 100,000 \\
\hline
\end{tabular}
\end{table}

To attain this goal, we apply the Kernel Farthest-First (KFF) algorithm \cite{Baram2012Online}, a greedy heuristic, to obtain exemplars from the malicious side of the SVM hyperplane.
KFF always chooses from the unlabeled query set the query that is the farthest from all the malicious queries in the labeled set.
Given unlabeled query set $U$ and labeled query set $L$, the farthest distance is defined as the maximum of the sum of distances between each malicious query $y$ in $L$ and a query $x$ in $U$ in kernel feature space.
Formally, the optimal $x$ is determined by
\begin{align}\label{e12}
\arg \max _{x\in U} < \sum _{x\in L} \left| \varphi \left( x\right) -\varphi \left( y\right) \right|>,
\end{align}

\noindent where $|\varphi(x)-\varphi(y)|$ is computed using Equation (\ref{e11}).
Each optimal $x$ is an exemplar, and is added to the labeled set.
As can be seen from Figure \ref{fig:SVMHYBRID}, ES obtains four exemplars, represented by four points surrounded by red circles.
One of the exemplars is a misclassified normal query.
This is unsurprising, since misclassification might happen.
Nevertheless, most exemplars are still truly malicious.
At each iteration of the adaptive detection process, suspicions and exemplars are manually labeled and used for refining the decision boundary of the meta classifier SVM of the detection model.

Specifically, SS uses the kernel distance to the hyperplane, which is already solved when a sample is classified by a SVM, while ES uses the kernel distance to malicious queries in current training set, which could be computed using the efficient-to-compute kernel function.
Thus SVM HYBRID is applicable for real-time attack detection scenarios.

\subsection{Training and classification}\label{Section3.4.training}

In machine learning, learning bias refers to any preference for choosing one hypothesis that explains the data over other equally acceptable hypotheses \cite{Sebastiani2002Machine}.
Meta-learning is the ability to learn from different learning biases that are appropriate for a particular problem by enriching the model hypothesis space \cite{Rendell1987Layered}.
Meta-learning is also an ensemble classifier technique \cite{rokach2010ensemble}.
Ensemble classifier techniques proved to be effective in intrusion detection, such as HMMPayl \cite{Ariu2011HMMPayl} and McPAD \cite{Perdisci2009McPAD}.
The ability to learn from different learning biases makes meta-learning an appropriate approach for malicious query detection, due to the complex, dynamic, confrontational nature of its problem domain.

In our approach, we employ stacked generalization \cite{Wolpert1992Stacked}, as well as named stacking, a well-known meta-learning technique, as our underlying detection model.
The key idea of stacking is to train a meta classifier from the prediction results of base classifiers \cite{Wolpert1992Stacked}.
By using a meta classifier, stacking tries to induce which base classifiers are reliable and which are not during training.
During classification, a test sample is classified by each of the base classifiers; then these classifications are fed into the meta classifier that makes the final decision.
The base classifiers in our stacking detection model are chosen from three different classifier families.
This combination scheme of base classifiers serves to obtain learning diversity and overcomes the learning bias of similar classifiers in the same category.
The meta classifier in our stacking detection model is SVM so that SVM HYBRID can operate on our stacking detection model by exerting its meta classifier SVM.

\bibliographystyle{plain}

\bibliography{mybibfile}


\section{Data collection and preprocessing}\label{dataset}

\subsection{Data collection}\label{Section4.1.dataset}

Web server logs are provided by most real-world web servers.
The availability of this data source is a great advantage of our analysis.
The web server logs of an academic institute during the period of July 1-10, 2016 are

used to evaluate our system.
The institute uses an Apache web server, whose logs are in CLF format as shown in

Figure \ref{fig:log}.
The web application chosen from the institute's server offers several resources on

its portal and consists of both static and dynamic pages.
In addition, this web application is equipped with a commercial WAF, called Hillstone

\cite{Hillstone}, and only requests passing through the WAF are recorded in its web

server access logs.

A description of the raw logs and data preprocessing procedure is presented in Table

\ref{tab:t2.dataPreprocess}.
The number of original requests, queries after data cleaning, data normalization and

application of character filter are shown.
The remaining 112,397 queries are used for our analysis.

The remaining queries are labeled by means of the signature set of ModSecurity \cite

{ModSecurity} and manual analysis, as described in Section \ref{Section5.discussion}.
ModSecurity is an open-source WAF, used to protect Apache web servers from web

attacks.
The signatures of ModSecurity are provided by the OWASP ModSecurity Core Rule Set

(CRS) \cite{CRS} project; the CRS version we use is 2.2.9.
The reason of using ModSecurity signatures to identify attacks is its relatively

better detection performance compared with other WAFs \cite{Prandl2015A}.
It is worth noting that we only include the malicious queries of the top four most

threatening attacks in our dataset, namely XSS, SQLI, RFI and DT attacks, as noted in

Section \ref{Section2.backRela}.

Web application concept drift occurred on the academic website within reason during

the ten-day experimental period, as described in Section \ref{Section5.discussion}.
In order to accurately evaluate the performance of our adaptive detection method, we

need to alleviate the effect of concept drift on our system.
Therefore, normal queries during the ten days are mixed up and then equally divided

into ten normal query sets.

Eleven disjointed query sets are used in our experiments, including an initial query

set and ten unknown query sets.
The initial set is used to induce the initial detection model, while each unknown set

is used to update the detection model.
The eleven sets are created from the entire 112,397 queries during the ten days.
Each unknown query set is composed of 10,000 queries on one day sequentially, and

12,397 queries are left.
The number of malicious queries in the ten unknown query sets is imbalanced: 95, 112,

97, 100, 107, 96, 92, 99, 127 and 108.
To make the number of malicious queries in each unknown set equal, we use the minimum

number of 92 on the $7^{th}$ day as the number of malicious queries in each unknown

set.
Then we replace surplus malicious queries in other nine unknown sets with normal ones

in the remaining 12,397 queries.
These queries are then used to create the initial query set.
The detailed statistics of the eleven sets of queries are presented in Table \ref

{tab:t3.dataset_statistics}.

\subsection{Data preprocessing}\label{Section3.3.Datapreparation}

\subsubsection{Data preparation}\label{Section3.3.1}

Our system models queries in HTTP GET requests shown in Figure \ref{fig:log}.
To collect our query dataset, the following steps are employed.

\emph{Data cleaning}.
Web server logs are examined to collect successful GET requests: requests whose

return code is equal to or greater than 200, and less than 300.
Then, static requests (e.g., .html, .wav, .txt, .jpg) are removed.
Finally, the remaining successful GET requests are parsed to extract queries like

$parameter1=value1\&parameter2=value2$ in Figure \ref{fig:log}.

\emph{Data normalization}.
Data normalization helps tighten the input space, including decoding printable ASCII

characters, un-escaping, transforming to lowercase and removing queries whose length

is less than four.
After normalization, initially different queries may appear identical.
The identical ones show as one entry in our dataset.

\emph{Character filter}.
The general syntax of URI in RFC 2616 \cite{Fielding1999RFC} defines the unsafe

characters that should not appear in standard queries.
We consider queries that contain any of the unsafe characters to be malicious and

propose a character filter to remove these queries.
The unsafe characters defined in RFC 2616 are as follows:
\begin{enumerate}
\item Any ASCII control character (octets 0-31, 128-255) and DEL (127)
\item ASCII SP (32)
\item ", \#, \%, $<$, $>$
\end{enumerate}

\subsubsection{Feature construction}\label{Section3.3.2}

Feature construction transforms each query into a numerical feature vector of $N$-

grams.
An $N$-gram is an $N$-character slice of a string \cite{cavnar1994n}, and its

analysis has been successfully applied to intrusion detection \cite

{Wang2004Anomalous,Wang2006Anagram,Jamdagni2013Review,Oza2014HTTP}.
We use $N$-gram analysis in feature extraction, taking advantage of its two

properties.
$N$-gram analysis captures the character distribution and sequence characteristics of

a string, and using these, it is able to distinguish malicious queries from normal

ones.
Furthermore, $N$-gram analysis is fully automatic and requires no prior knowledge

about target web application and target attacks.

The value of $N$ is important in $N$-gram analysis.
When $N$-gram ($N\geq$1) features are extracted, the input feature space of labeled

query set is defined as
\begin{equation}\label{e13}
S=\left\{ N\raisebox{0mm}{-}gram_{i}|i=1,\ldots ,63^{N}\right\}.
\end{equation}

The size of $S$ is $63^N$, since 63 unique characters exist in all the queries after

data preparation.
Intuitively, sufficiently high-order $N$-grams must be used to obtain more accurate

features.
However, for $N\raisebox{0mm}=\raisebox{0mm}1$, $N$-gram features only capture the

character distribution information and lacks the sequence information, which is

necessary for attack detection in strings.
For $N\raisebox{0mm}=\raisebox{0mm}2$, the character sequence information is also

included.
For $N\raisebox{0mm} \textgreater\raisebox{0mm}2$, the number of $N$-grams

exponentially increases with $N$.
The smallest number of features is already 25,507 ($63^3$), which exacts a high cost

in time and computational complexity.
Therefore, we choose $N\raisebox{0mm}=\raisebox{0mm}2$ to balance computational

consumption and detection performance, with only 3,969 ($63^2$) distinct $N$-grams

used.
65,536 ($256^2$) $N$-grams would be involved in the feature space without data

normalization and character filter.
It is observable that the size of feature space decreases by one order of magnitude

after the utilization of data normalization and character filter.

Two varieties of $N$-gram models are binary-based and frequency-based.
To capture the structure of queries more accurately, a frequency-based $N$-gram model

is applied.
Each $N$-gram frequency is then divided by the most frequently appearing $N$-gram

frequency in the same query.

\subsubsection{Feature reduction}\label{Section3.3.3}

Feature selection serves to remove from the feature set the $N$-grams that do not

appear in all the queries, and identify $N$-grams that aid classification.

The number of useful $N$-grams obtained by feature selection is around 1,000, which

is still a large number.
A large feature set is impractical in our learning task, since our query dataset is

huge.
In addition, vast features usually do not enhance classification performance and

sometimes even degrade it.
Hence, to alleviate the curse of dimensionality, dimensionality reduction is

implemented after feature selection.

Juvonen et al. \cite{Juvonen2015Online} also applied dimensionality reduction to the

feature space of queries.
Our approach differs from their work in that we apply feature selection before

dimensionality reduction in order to reduce dimensionality reduction and training

time, as well as improve classification performance.

\bibliographystyle{plain}

\bibliography{mybibfile}


\section{Experiments and evaluation}\label{Section4.experiment}

In this section, we report and analyze the results of a series of experiments.
Experiment 1 (Section \ref{Section4.2.Experim1}) is conducted to determine the best

setting of feature reduction and classifiers used for our stacking-based detection

model.
Experiment 2 (Section \ref{Section4.3.Experim2}) observes the efficacy of SVM HYBRID

in AMODS, with respect to detection performance and the number of malicious queries

obtained, while AMODS is compared with constant models (SVM model and stacking model)

and other adaptive models (adaptive SVM model, SS, ES, SVM AL and random selection)

in Experiment 3 (Section \ref{Section4.4.Experim3}).
Lastly, Experiment 4 (Section \ref{Section4.4.Experim4}) compares AMODS with other

work re the detection performance on our dataset.

All the experiments are implemented in Java and conducted on a machine with 32-GB

memory and Inter{\textsuperscript{\textregistered}} Core{\textsuperscript

{\texttrademark}}  2.93-GHz CPU.
To allow SVM HYBRID to work well with the meta classifier SVM of the detection model,

we implement stacking on our own.
Other classifiers are implemented using WEKA \cite{witten1999weka}, except for SVM,

which is implemented using LibSVM \cite{Chang2007LIBSVM}.

To evaluate the classification performance of our detection model, the primary metric

employed is F-value, an effective measure for the data imbalance problem.
As defined in (\ref{e14}), F-value is a combination of Precision and Recall.
Precision measures the percentage of correctly detected malicious queries in all

detected malicious queries, while Recall shows the percentage of correctly detected

malicious queries in all truly malicious queries.
$\beta$ corresponds to the relative importance of Precision versus Recall and is

usually set to 1, as in \cite{Jamdagni2013Review}.
\begin{multline}\label{e14}
F-value=\left(1+\beta \right) ^{2}* \dfrac {Precision} {\beta ^{2}*(Precision

+Recall)}
\end{multline}

In order to comprehensively assess the performance of the detection model, True

Positive (TP) rate and FP rate are also utilized.

\subsection{Experiment 1: determination of the best setting of feature reduction and

stacked classifiers}\label{Section4.2.Experim1}

\begin{figure}[t!]
    \centering
    \includegraphics[width=0.7\columnwidth]{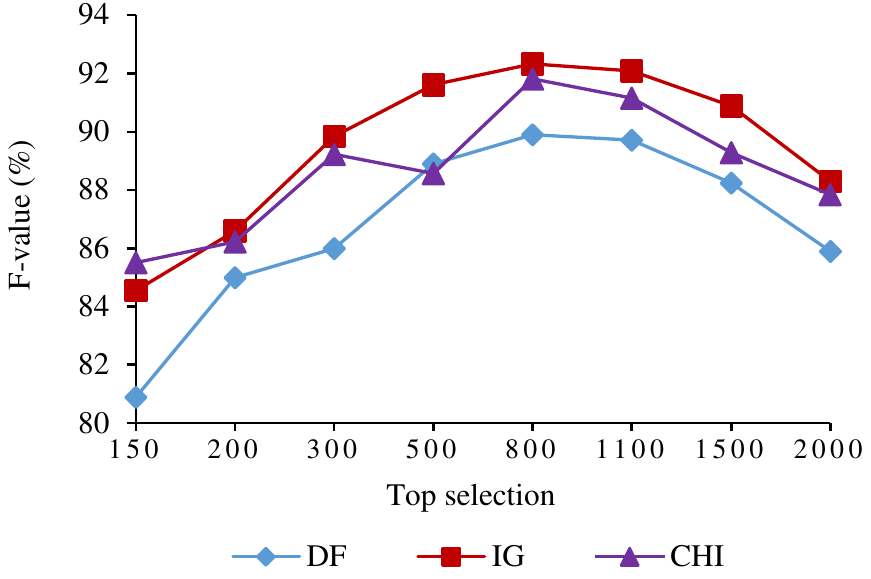}
    \caption{Comparisons of various combinations of feature selection methods and top

selection}
    \label{fig:featureSelection}
\end{figure}

Experiment 1 aims to determine the best setting of feature reduction and stacked

classifiers.
The superset of the eleven query sets shown in Table \ref{tab:t3.dataset_statistics}

is used.
The superset consists of 100,100 queries.
Queries in the superset are represented by $N$-gram feature vectors in the 3,969-

dimensional feature space introduced in Section \ref{Section3.3.2}.
The best combination of feature selection method and top selection, dimensionality

reduction method, and reservation quantity is separately determined on two SVMs, one

of which uses RBF kernel, and the other uses Polynomial kernel.
Additionally, base classifiers and meta classifier SVM (RBF and Polynomial kernel) of

our stacking-based detection model are also determined.
These experiments are implemented on the query superset using 10-fold cross

validation, amounting to 768 runs in total.
The determination metric is the average F-value of each setting.
Note that even though the best setting of feature reduction methods and stacked

classifiers may vary among different query datasets, the determination scheme can be

universally applied to them.

\begin{figure}[t!]
    \centering
    \includegraphics[width=0.7\columnwidth]{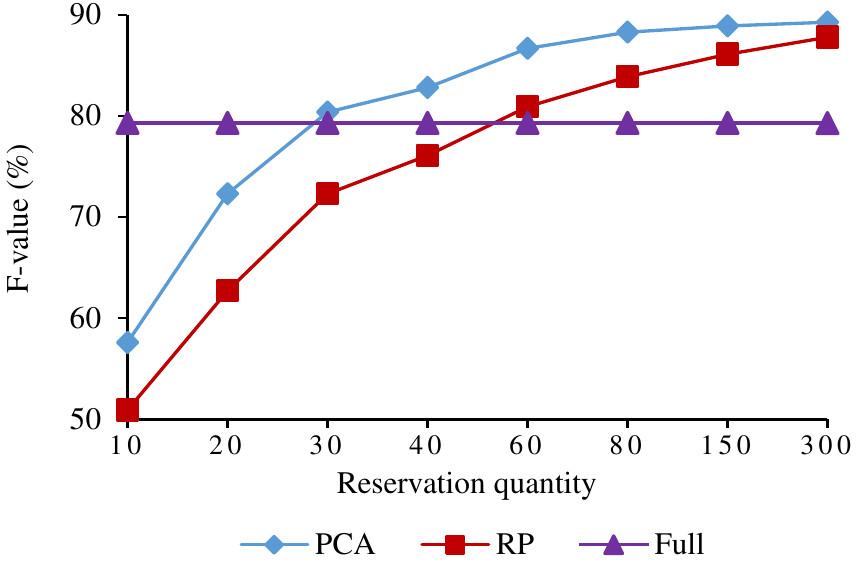}
    \caption{Comparisons of various combinations of dimensionality reduction methods

and reservation quantity}
    \label{fig:dimensionReduction}
\end{figure}

\emph{Feature selection and top selection.}
Various combinations of feature selection methods (Chi-square test, Information Gain

(IG) or Document Frequency (DF)) and top selection (150, 200, 300, 500, 800, 1100,

1500 or 2000) are compared in Figure ~\ref{fig:featureSelection}.
It shows that among the three feature selection methods, IG performs best.
IG and top 800 overwhelm other combinations, and thus are chosen as the best setting.

\emph{Dimensionality reduction and reservation quantity.}
Different combinations of dimensionality reduction methods (Principal Component

analysis [PCA] or Random Projection [RP]) and reservation quantities (10, 20, 30, 40,

60, 80, 150 or 300), as well as a full feature set, are compared in Figure \ref

{fig:dimensionReduction}.
As demonstrated, PCA outperforms RP at lower dimensions, and both of them gradually

stabilize when dimensions increase.
PCA surpasses the full feature set at the reservation quantity of 80.
PCA and 80 reserved dimensionalities are determined as the optimal setting.

\begin{table*}[t!]
\centering
\caption{Determining base classifiers and meta classifier for stacking-based

detection model}
\label{tab:t5.all classifiers}
\begin{tabular}{llcccc} \hline
\multirow{2}{*}{Classifier type} &\multirow{2}{*}{Classifiers}  & F-value on & F-

value & TP rate & FP rate \\
\hhline{~~~---}   & & training set(\%) & \multicolumn{3}{c}{In 10-fold cross

validation (\%)} \\
\hline
& MLP & 85.61 & 83.77 & 86.96 & 0.19 \\
& RBF networks & 31.54 & 23.54 & 98.91 & 5.95 \\
& BayesNet & 99.87 & 62.90 & 96.74 & 1.03 \\
& Naive Bayes & 100.00 & 60.14 & 96.74 & 1.16 \\
& Random Forest & 93.17 & 9192 & 98.90 & 0.15 \\
Single classifiers& J48graft & 92.65 & 91.28 & 96.73 & 0.23 \\
& LMT & 91.50 & 90.91 & 97.83 & 0.16 \\
& JRip & 84.10 & 81.08 & 97.89 & 0.40 \\
& Decision table & 57.43 & 52.49 & 85.77 & 1.31 \\
& Conjunctive rule & 59.98 & 54.37 & 97.83 & 1.50 \\
& SimpleLogistic & 92.92 & 90.10 & 98.91 & 0.19 \\
& Logistic & 77.94 & 76.50 & 76.09 & 0.21 \\
\hhline{------}
& RF SL MLP & 96.26 & 94.30 & 98.91 & 0.10 \\
& RF SL JRip & 93.87 & 93.26 & 97.83 & 0.11 \\
Base classifiers& RF JRip MLP & 92.98 & 89.11 & 97.83 & 0.20 \\
& MLP SL JRip & 92.13 & 90.09 & 98.91 & 0.19 \\
& MLP SL JRip RF & 94.37 & 92.32 & 98.24 & 0.16 \\
\hhline{------}
\multirow{2}{*}{Meta classifiers}& SVM-RBF & 94.21 & 93.23 & 98.99 & 0.10 \\
& SVM-Poly & 93.63 & 91.19 & 97.96 & 0.19 \\
\hline
\end{tabular}
\end{table*}

\emph{Classifier selection.}
Using the best setting of feature selection and dimensionality reduction, classifier

selection is carried out in three steps as shown in Table \ref{tab:t5.all

classifiers}.
(a) \emph{Determine base classifier candidates.}
Twelve single classifiers are tested, which are chosen from five classifier families

\cite{han2011data}: neural networks, Bayesian, tree based, rule based, and regression

based classifiers.
It can be observed from Table \ref{tab:t5.all classifiers} that the performance of

single classifiers in different families varies greatly.
Even though Bayesian classifiers perform well in the case of the training set, they

achieve low F-values of about 60\% in the case of cross validation.
MLP, Random Forest, JRip and SimpleLogistic perform best in their respective family

in both cases.
Hence, these four classifiers are chosen as base classifier candidates.
(b) \emph{Determine the best combination of base classifiers.}
All combinations of three base classifier candidates, together with the combination

of all the base classifier candidates are tested in stacking with the meta classifier

being SVM with RBF kernel and Polynomial kernel respectively.
Table \ref{tab:t5.all classifiers} indicates that the combination of Random Forest,

Logistic and MLP outperforms not only other combinations but also the best single

classifier, namely, Random Forest.
Therefore, this combination of base classifiers is used in our stacking-based

detection model.
(c) \emph{Determine the best meta classifier SVM.}
Table \ref{tab:t5.all classifiers} shows RBF kernel outperforms the Polynomial

kernel, thus SVM with RBF kernel is determined as the meta classifier of the

detection model.
The optimal pair of SVM penalty parameter $C$ and RBF kernel parameter $\gamma$ is

$(C,\gamma)=(0.05,2)$, obtained using grid-search cross-validation \cite

{hsu2003practical}.

\subsection{Experiment 2: evaluation of AMODS}\label{Section4.3.Experim2}
\begin{table}[t!]
\centering
\caption{SVM HYBRID parameters}
\label{tab:t7.parameters}
\begin{tabular}{lcccccccc} \hline
\multirow{2}{*}{Parameter}   & \multicolumn{5}{c}{Adjustable}  & \hspace{0.1cm} &

\multicolumn{2}{c}{Fixed}\\
\hhline{~-----~--}    & \multicolumn{5}{c}{$\theta$} &\hspace{0.1cm} & $M$ & $R$ \\
\hline
Case & A & B & C & D & E &\hspace{0.1cm} & A-E & A-E \\
\hhline{~~~~~~~~~}
Value & 5:5 & 6:4 & 7:3 & 8:2 & 9:1 &\hspace{0.1cm} & 150 & 5\\
\hline
\end{tabular}
\end{table}

Experiment 2 reports and discusses the efficacy of our system, specifically, the

impacts of SVM HYBRID on both the detection performance and the number of malicious

queries obtained, by adjusting $\theta$ (the ratio of the size of query subset for SS

to that for ES), which is crucial for SVM HYBRID.

The process introduced in Section \ref{section3.1.System} is carried out using

different values of $\theta$, including 1:9, 2:8, 3:7, 4:6, 5:5, 6:4, 7:3, 8:2 and

9:1.
In the following, we examine detailed performances of the last five cases, which are

denoted as A, B, C, D and E respectively, and whose parameters are listed in Table

\ref{tab:t7.parameters}.
Two fixed parameters are involved.
The number of obtained important queries (suspicions and exemplars) on each day,

denoted as $M$ $-$ which should be larger than the number of malicious queries in

each unknown set (92) $-$ is set to 150.
In addition, the average size of clusters in $K$-medoids clustering in SS, denoted as

$R$, is set to 5, a quantity we consider appropriate.
Then the number of clusters, namely $K$, is solved as the largest integer value of

the number of unknown queries in the confusing region divided by $R$.


\begin{table}[t]
\centering
\caption{Comparisons of ratios ($\mu$)}
\label{tab:t8.ratio}
\begin{tabular}{clllll} \hline
\multirow{2}{*}{Day}  & \multicolumn{5}{c}{$\mu$ for various cases} \\
\hhline{~-----}    & A & B & C & D & E \\
\hline

1 & 2.0 & 1.5 & 1.1 & 0.9 & 0.7 \\
2 & 3.4 & 2.8 & 2.2 & 1.8 & 1.5 \\
3 & 6.1 & 5.0 & 4.0 & 3.4 & 3.0 \\
4 & 11.5 & 9.0 & 7.8 & 6.9 & 5.8 \\
5 & 24.0 & 20.4 & 17.8 & 15.7 & 14.0 \\
6 & 49.0 & 36.5 & 29.0 & 29.0 & 24.0 \\
7 & 74.0 & 49.0 & 74.0 & 36.5 & 36.5 \\
8 & 74.0 & 74.0 & 49.0 & 49.0 & 49.0 \\
9 & 149.0 & 74.0 & 74.0 & 74.0 & 49.0 \\
10 & 149.0 & 149.0 & 149.0 & 74.0 & 74.0 \\

\hline
\end{tabular}
\end{table}

\begin{figure}[t!]
    \centering
    \includegraphics[width=0.7\columnwidth]{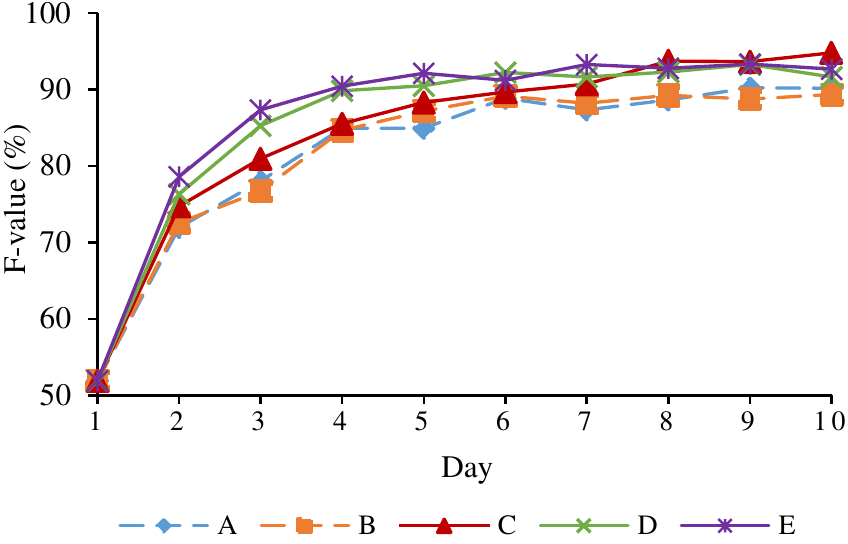}
    \caption{Comparisons of F-values}
    \label{fig:F-value.F7}
\end{figure}

\begin{figure}[t!]
    \centering
    \includegraphics[width=0.7\columnwidth]{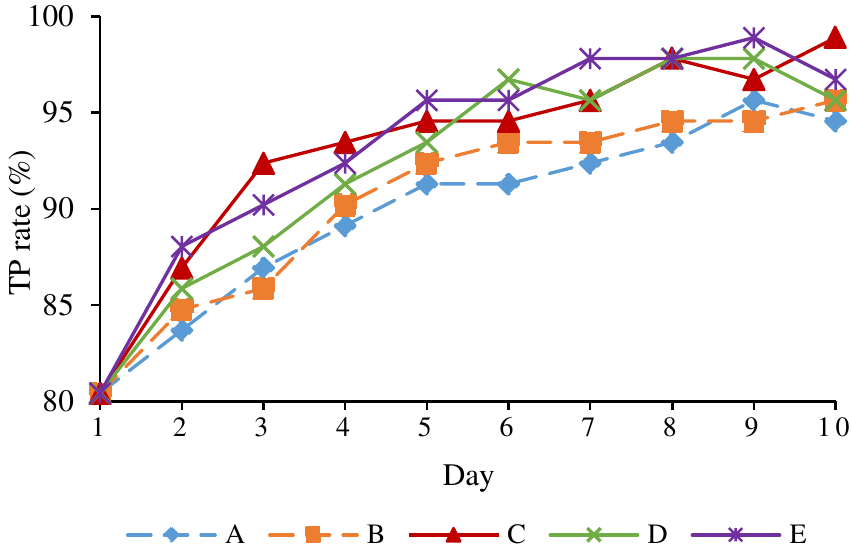}
    \caption{Comparisons of TP rates}
    \label{fig:TPrate.F8}
\end{figure}

\begin{figure}[t!]
    \centering
    \includegraphics[width=0.7\columnwidth]{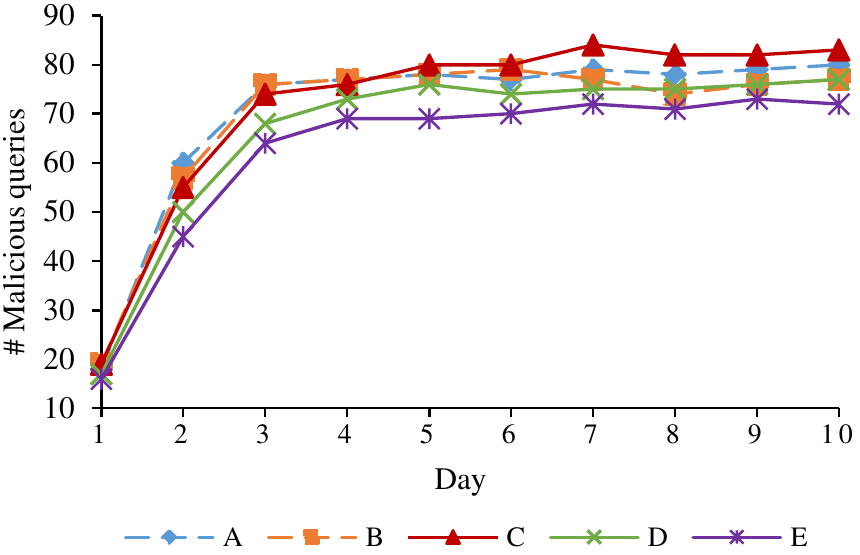}
    \caption{\#Malicious queries obtained}
    \label{fig:f9.maliciousqueries}
\end{figure}

The ratio of exemplar quantity to suspicion quantity is denoted as $\mu$.
Values of $\mu$ of the five cases during the ten-day detection are shown in Table

\ref{tab:t8.ratio}.
As is evident from Table \ref{tab:t8.ratio}, all cases show an increasing trend.
This phenomenon is due to the decreasing number of suspicions over time.
Exemplars are incremented to make up the reduction in suspicions, so that the total

of suspicions and exemplars can stay constant.
It can be concluded that ES becomes increasingly dominant in SVM HYBRID during the

adaptive detection process.

F-values of the five cases are compared in Figure \ref{fig:F-value.F7}, which shows

that all the cases present an increasing trend.
For the first three days, relatively low F-values are shown for all cases.
At the beginning only a few queries are in the training pool, so the decision

boundary of the detection model tends to be inaccurate.
With more suspicions and exemplars incorporated into the training pool, the detection

performance is rapidly improved.
During the ten days, the trends of cases C, D and E always appear above that of A and

B.
Recall that $\theta$s of C, D and E are larger than that of A and B, as shown in

Table \ref{tab:t7.parameters}; thus C, D and E obtain more suspicions.
It can be inferred that suspicions contribute more to detection performance in

comparison with exemplars.
This is reasonable because the updated detection model would make more accurate

prediction of new queries, when the most important uncertain (informative) queries

(suspicions) are incorporated into the training pool, rather than representative

certain queries (exemplars).

\begin{table}[!t]
\centering
\caption{Comparisons of FP rates}
\label{tab:t9.fp-rates}
\begin{tabular}{cccccc} \hline
\multirow{2}{*}{Day}  & \multicolumn{5}{c}{FP rates (\%) for various cases} \\
\hhline{~-----}    & A & B & C & D & E \\
\hline

1 & 1.20 & 1.20 & 1.20 & 1.20 & 1.20  \\
2 & 0.45 & 0.45 & 0.42 & 0.36 & 0.33 \\
3 & 0.33 & 0.35 & 0.33 & 0.17 & 0.15 \\
4 & 0.19 & 0.21 & 0.23 & 0.11 & 0.11 \\
5 & 0.19 & 0.18 & 0.18 & 0.12 & 0.11 \\
6 & 0.13 & 0.15 & 0.15 & 0.12 & 0.13 \\
7 & 0.15 & 0.17 & 0.14 & 0.12 & 0.11 \\
8 & 0.16 & 0.16 & 0.10 & 0.13 & 0.12 \\
9 & 0.15 & 0.17 & 0.09 & 0.11 & 0.12 \\
10 & 0.17 & 0.17 & 0.09 & 0.12 & 0.11 \\

\hline
\end{tabular}
\end{table}

As for cases C, D and E, the trend of case C is naturally expected to be below those

of D and E, since more suspicions mean better detection performance, as just

inferred.
However, case C reverses this expectation: its increasing trend is slightly steeper

than those of D and E, and exceeds them on the $8^{th}$ day.
A closer inspection of the three cases reveals that case C obtains the largest total

number of misclassified suspicions before the $8^{th}$ day among the three cases.
We attribute the superiority of case C to the misclassified suspicions it obtains,

because misclassified suspicions are more informative than other correctly-classified

queries.
As a result, incorporating them contributes to a more accurate decision boundary of

the detection model.
On the last day, case C achieves the highest F-value (94.79\%) among the five cases

during the ten days, which implies the potential of case C to improve detection

performance.

\begin{table}
\centering
\caption{The process of case C}
\label{tab:t10.caseC}
\begin{tabular}{ccccc} \hline

\multirow{2}{*}{Day}  & \multicolumn{4}{c}{Number of queries} \\
\hhline{~----}     & Margin & Conf. & Suspicions & Exemplars \\
\hline

1 & 450 & 351 & 70 & 80 \\
2 & 339 & 237 & 47 & 103 \\
3 & 257 & 151 & 29 & 121 \\
4 & 203 & 87 & 17 & 133 \\
5 & 152 & 41 & 8 & 142 \\
6 & 131 & 24 & 4 & 146 \\
7 & 114 & 16 & 3 & 149 \\
8 & 98 & 14 & 2 & 148 \\
9 & 85 & 10 & 2 & 148 \\
10 & 59 & 8 & 1 & 149 \\

\hline
\end{tabular}
\end{table}

Figure \ref{fig:TPrate.F8} and Table \ref{tab:t9.fp-rates} exhibit TP rates and FP

rates of the five cases respectively.
Figure \ref{fig:TPrate.F8} explicitly shows that their TP rates all gradually

increase as days go on, while Table \ref{tab:t9.fp-rates} suggests that their FP

rates rapidly decline for the first three days and then gradually stabilize for the

duration.
On the $10^{th}$ day, the TP rates and FP rates of the five cases fall in the ranges

of [95.65\%,98.91\%] and [0.09\%,0.17\%] separately.
On this day, case C achieves the highest TP rate and lowest FP rate of the ten days

among all the cases.

\begin{figure}[t!]
    \centering
    \includegraphics[width=0.7\columnwidth]{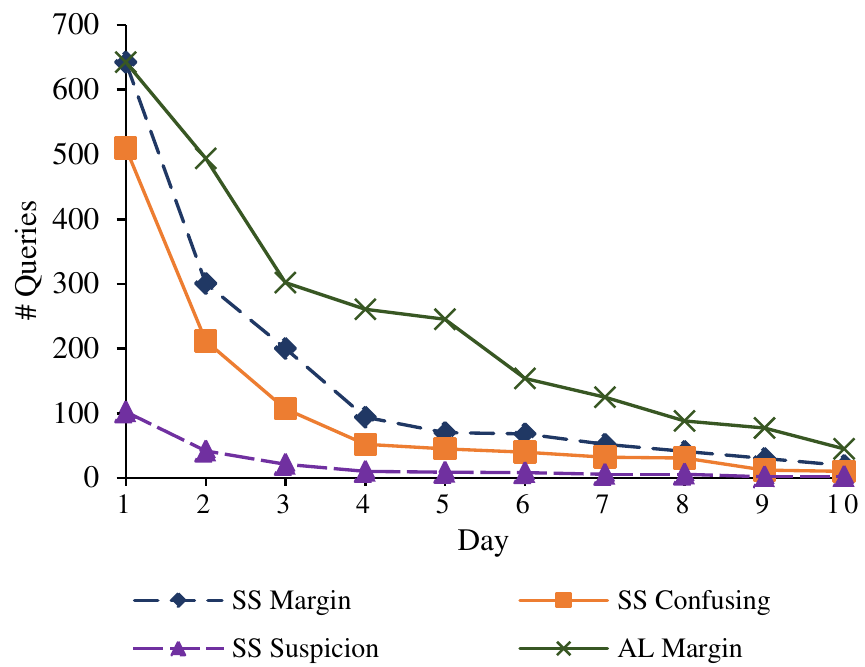}
    \caption{SS vs. SVM AL}
    \label{fig:f10.processofss}
\end{figure}

\begin{figure}[t!]
    \centering
    \includegraphics[width=0.7\columnwidth]{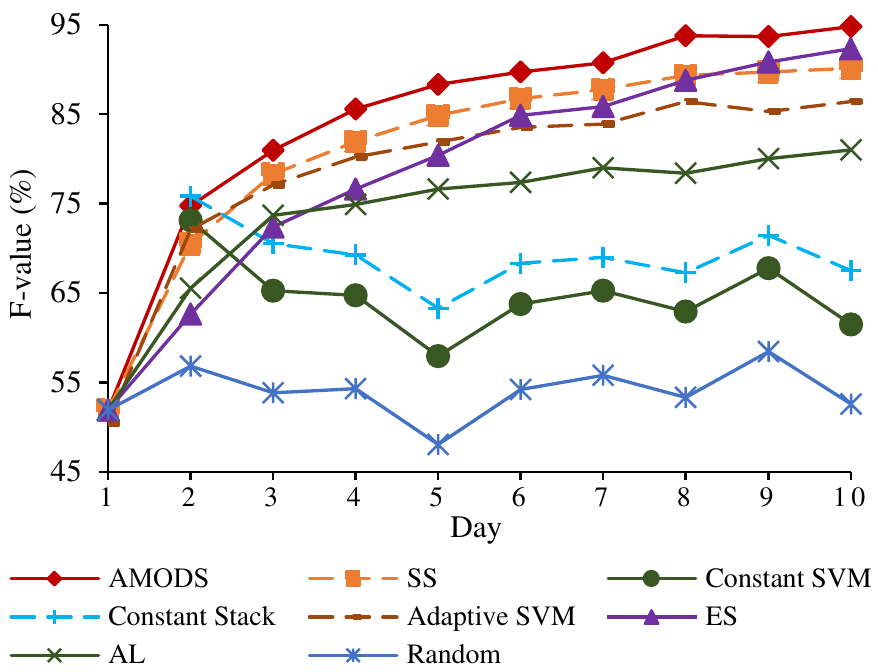}
    \caption{Comparisons of F-values}
    \label{fig:f11.fvaluesoffive}
\end{figure}

\begin{figure}[t!]
    \centering
    \includegraphics[width=0.7\columnwidth]{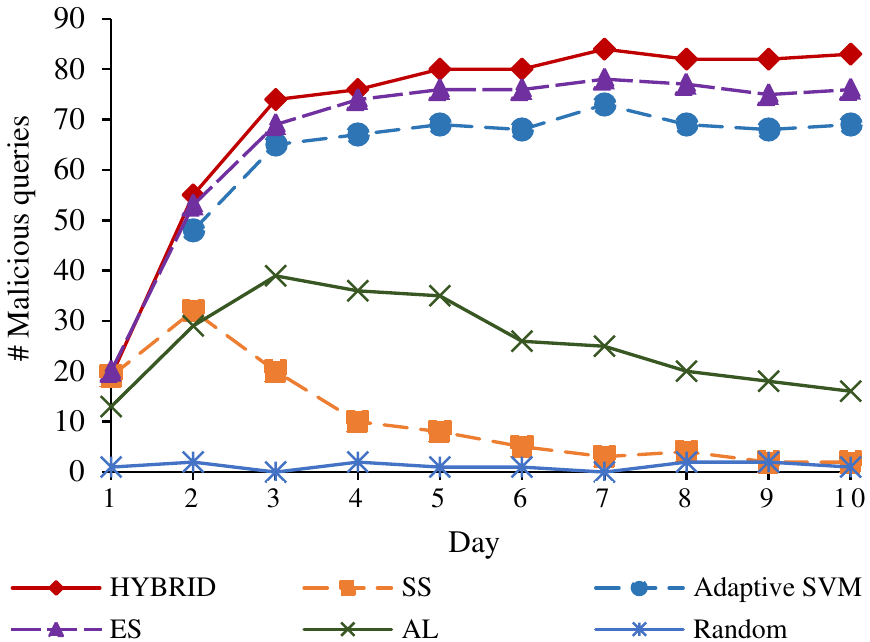}
    \caption{\# Malicious queries obtained}
    \label{fig:f12.numberMalicious}
\end{figure}

As analyzed above, case C outperforms other cases.
Since $\theta$ is 7:3 in case C, the each-day unknown subset for SS and ES in case C

consists of 7,000 queries and 3,000 queries respectively.
Table \ref{tab:t10.caseC} shows the process of case C during the ten-day detection.
As Table \ref{tab:t10.caseC} shows, on the first day, SVM HYBRID starts with SS: the

SVM margin covers 450 queries in the unknown subset of SS, of which 351 queries fall

into the confusing region.
Then, $K$-medoids clustering is performed on the 351 queries and obtains 70

suspicions.
Afterwards, SVM HYBRID switches to ES on the unknown subset for ES and obtains 80

exemplars.
Obtained suspicions and exemplars are manually labeled, and added to the training

pool to update the detection model.
This process repeats for the remaining days.

Truly malicious queries obtained by our system can be used to update the WAF

signature library.
Hence, we expect that our system could maintain a desirable detection performance and

simultaneously obtain as many malicious queries as possible.
Figure \ref{fig:f9.maliciousqueries} presents the trends of the number of malicious

queries obtained by the five cases.
It is clear that they all show an increasing trend.
This is attributable to the fact that the number of exemplars increase while

suspicions decline during the adaptive detection process, as concluded from Table

\ref{tab:t8.ratio}.
Suspicions are informative queries, yet exemplars are representative malicious

queries.
Hence, since exemplars dominate the total malicious queries obtained, the increase in

the number of exemplars results in the increase in the number of total malicious

queries.
Last but not least, case C obtains the largest number of malicious queries among the

five cases, not only on the last day but for the entire ten days.
This confirms that suspicions help improve both the detection ability and the number

of correctly classified exemplars, especially when $\theta$ is close to 7:3.

In summary, all the cases of AMODS yield desirable performance.
The experiment results demonstrate the contribution of SVM HYBRID to AMODS: SS mainly

helps improve the detection performance, while ES mainly helps obtain malicious

queries.
Since the ultimate goal of our system is to achieve a good detection performance and

simultaneously obtain as many unknown malicious queries as possible, case C is chosen

as the optimal case and 7:3 as the optimal value for $\theta$, which slightly

outperforms other cases in the two respects.

\subsection{Experiment 3: comparisons of constant models and adaptive models}\label

{Section4.4.Experim3}

Experiment 3 aims to compare AMODS (case C) with constant models and adaptive models.
Constant models are trained only once on the superset of the initial query set and

the first-day unknown query set, including the constant SVM model and the constant

stacking model.
Five adaptive models are also tested: an adaptive model combining both stacking and

SVM HYBRID in our proposed AMODS, an adaptive SVM model, SS, ES, random selection,

and SVM AL, which is an existing effective adaptive learning method (see Section

\ref{Section3.2.1.SS}).
All the models in Experiment 3 are implemented in AMODS to carry out the ten-day

detection with same predefined parameters as in Experiment 2, except that SS and SVM

AL are required to obtain \emph{at most} 150 important queries on each day, namely,

$M$$\leq$150.

\begin{table*}[t]
\centering
\caption{Comparisons of web attack detection methods}
\label{tab:detectionmethods}
\begin{tabular}{lccc} \hline
Method & F-value(\%) & TP rate (\%) & FP rate (\%) \\
\hline
Linear combination \cite{Kruegel2003Anomaly} & \textbf{71.73} & \textbf{92.39} & 0.61

\\
Wavelet transform \cite{Kozakevicius2015URL} & 61.04 & 82.61 & 0.81 \\
Dimensionality reduction \cite{Juvonen2015Online} & 70.48 & 86.96 & \textbf{0.56} \\
Adaptive learning (AMODS)& \textbf{94.79} & \textbf{98.91} & \textbf{0.09} \\
\hline
\end{tabular}
\end{table*}

Figure \ref{fig:f10.processofss} depicts the process of SS in regard to the number of

queries within the SVM margin (SS Margin), within the confusing region (SS

Confusing), and obtained as suspicions (SS Suspicion) separately.
SVM AL is also shown about the number of queries within the margin (AL Margin).
They all present a decreasing trend.
In comparison with SVM AL, important queries obtained by SS (SS Suspicion) are

constantly fewer than those obtained by SVM AL (the smaller of AL Margin and 150).
However, queries within the margin in SS are always fewer than those in SVM AL, which

means the uncertainty of the detection model in SS is smaller than that in SVM AL.
This phenomenon implies that SS is superior to SVM AL in reducing classifier

uncertainty.

F-values of the eight detection approaches are exhibited in Figure \ref

{fig:f11.fvaluesoffive}.
Among the five methods, random selection shows the worst performance and constantly

fluctuates.
The failure of random selection is attributed to the trivial probability of a

randomly selected query being informative or representative.
The constant SVM model and the constant stacking model perform better only than

random selection; similar trends of the three methods also indicate that randomly

selected queries make little contribution to detection performance improvement.
Yet, for the five adaptive methods, steep uptrend is shown for the first three days

and then steady uptrend for the remaining days.
Detection performance improvement of AMODS over constant methods suggests that AMODS

is able to effectively leverage existing knowledge, including knowledge gained during

the adaptive detection process, toward enhanced
subsequent detection of queries.
Moreover, AMODS along with SS outperforms other adaptive methods.
The improvement of AMODS over the adaptive SVM model implies the stronger

classification ability of stacking over single classifiers.
The superiority of AMODS mainly lies in that it employs SVM HYBRID, which combines

the contributions of SS and ES while also mitigating their limitations.
Both SS and ES boost the performance of AMODS.
Furthermore, SS outperforms ES, which demostrates that suspicions make a valuable

contribution to an accurate detection model, as validated in Figure \ref{fig:F-

value.F7}.
SS also outperforms SVM AL, even though SS obtains fewer important queries, as shown

in Figure \ref{fig:f10.processofss}.
This phenomenon reveals the advantage of SS over SVM AL: SS requires less manual

labeling work and achieves better detection performance.
The main weakness of SVM AL is its ignorance of density distribution of unknown

samples, which leads to sample bias, while SS preserves the density distribution by

using $K$-medoids clustering to obtain suspicions.

Figure \ref{fig:f12.numberMalicious} compares the number of truly malicious queries

obtained by the six adaptive detection methods.
Random selection still performs the worst.
AMODS, together with ES and the adaptive SVM model, outperforms other methods, with a

total number of 715, 656 and 615 malicious queries obtained respectively.
The adaptive SVM model lags behind among the top three methods, which demonstrates

the strong detection ability of the stacking-based detection model.
Moreover, the reason why ES falls short of AMODS lies in that SS benefits the

detection performance of AMODS, which leads AMODS to obtain more correctly-classified

exemplars than ES does.
Generally, AMODS, ES and the adaptive SVM model show an uptrend, while SS and SVM AL

show a downtrend.
This distinction is due to the difference in their preference for queries.
AMODS, ES and the adaptive SVM model prefer queries that are highly likely to be

malicious, while SS and SVM AL prefer queries that the classifier is uncertain about,

and thus might not be malicious.
The reduction in uncertain queries leads to a decline in malicious queries obtained

by SS and SVM AL.
During the ten-day detection, the total number of malicious queries obtained by AMODS

(715) is 2.78 times and 59.58 times that by SVM AL (257) and random selection (12)

respectively.
This demonstrates the overwhelming ability of AMODS in harvesting malicious queries.

In conclusion, Experiment 3 demonstrates that adaptive learning can further enhance

malicious query detection performance over constant models.
Experiment results reveal the advantages of AMODS, SS and ES over SVM AL, constant

models and random selection in two respects: improving detection performance and

obtaining more malicious queries.
AMODS wins in both respects among all the methods.
SS does well in the former respect, while ES excels in the latter.
Both SS and ES defeat other methods except for AMODS in their respective strengths.

\begin{figure}[t]
    \centering
    \includegraphics[width=0.9\columnwidth]{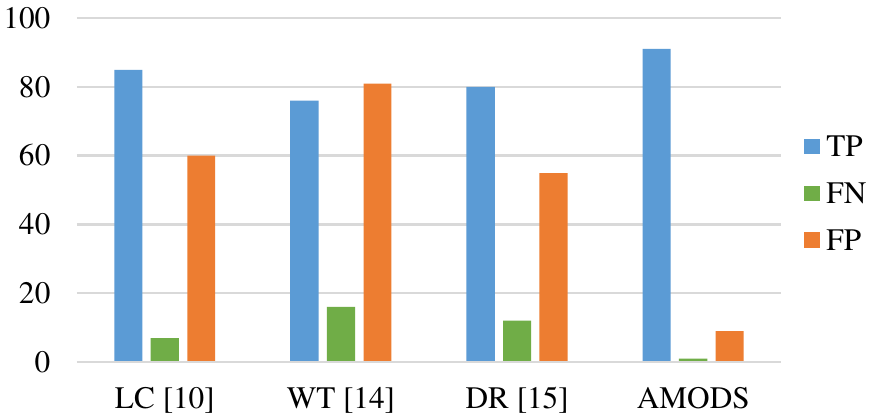}
    \caption{Comparisons of web attack detection methods}
    \label{fig:compareWeb}
\end{figure}

\subsection{Experiment 4: comparisons with related work}\label{Section4.4.Experim4}

In the context of web attack detection, existing methods are all constant and thus

cannot adapt to the change of malicious behaviors.
Our proposed adaptive detection method is capable of detecting the latest attacks by

incrementally updating the detection model.

Experiment 4 aims at comparing AMODS with existing detection methods.
However, our proposed method cannot be directly compared with them, since there is no

public data available for web attack detection and existing works all used their own

datasets for evaluation.
The only labeled, publicly available trace in intrusion detection is DARPA 1999

\cite{lippmann20001999}, whose HTTP traffic is not appropriate for web attack

detection, because it does not contain many web attacks \cite{Mchugh2000Testing}.
For this reason, we use our dataset to test existing web attack detection methods,

including linear combination \cite{Kruegel2003Anomaly}, wavelet transform \cite

{Kozakevicius2015URL} and dimensionality reduction \cite{Juvonen2015Online}.

We set the anomaly threshold of \cite{Kruegel2003Anomaly} to the default 10\% larger

than the maximum anomaly score seen during training.
\cite{Kruegel2003Anomaly} is conducted on our ten unknown sets within 10-fold cross

validation.
Since \cite{Kruegel2003Anomaly} uses normal traffic to build the detection model,

malicious queries in the training set of each fold are removed.
\cite{Kozakevicius2015URL} and \cite{Juvonen2015Online} are unsupervised, so do not

require a training phase.
We set their anomaly threshold to 0.92\%, the percentage of malicious queries in our

dataset.
TW2D-BA is used to perform \cite{Kozakevicius2015URL}, while PCA along with RM is

used to implement \cite{Juvonen2015Online}.
\cite{Kozakevicius2015URL} and \cite{Juvonen2015Online} are implemented on each

unknown dataset, and their performance is averaged separately.

Table \ref{tab:detectionmethods} and Figure \ref{fig:compareWeb} compare the

detection performance of AMODS with the three web attack detection methods.
The comparison results illustrate that our method is the overall winner and

significantly beats the other methods, achieving the highest F-value (94.79\%) and TP

rate (98.91\%), as well as the lowest FP rate (0.09\%) among them.
Among the three methods, \cite{Kruegel2003Anomaly} obtains the highest F-value

(71.73\%) and the highest TP rate (92.39\%).
However, they are still lower than the lowest F-value and TP rate obtained by AMODS

on the $10^{th}$ day, namely 91.67\% and 95.65\%.
Moreover, their lowest FP rate (0.56\%) is higher than the highest of SVM HYRBID on

the $10^{th}$ day (0.17\%).

We attribute the success of AMODS to the principle of SVM HYBRID, namely, choosing

both the most important prone-to-be-misclassified queries (suspicions) and the most

representative malicious queries (exemplars) from unknown queries.
By incrementally incorporating correctly-labeled suspicions and exemplars into the

training pool to update the detection model, its detection ability is greatly

enhanced.
The detection model is up-to-date and capable of detecting the most recent unknown

attacks.

\bibliographystyle{plain}

\bibliography{mybibfile}


\section{Discussion}\label{Section5.discussion}

In this paper, we propose an adaptive learning strategy called SVM HYBRID for the efficient selection of important queries, and develop a malicious query detection system  that can adaptively update the detection model with a small manual labeling workload.
In this section, we clarify some relevant issues of our approach and discuss the limitations of our approach.

\emph{Labeling unknown queries.}
Manually labeling the 112,397 queries in web requests is time-consuming and error-prone.
Therefore, we initially used the rules of ModSecurity to detect attacks automatically.
Then we manually confirmed the detected attacks, and concluded that the TP rate and FP rate of the detection results was 33.37\% and 8.64\%.
To decipher the reason for the unsatisfactory detection results, we manually checked the ModSecurity rule set and found that some rules matched queries with simple malicious code snippets, which might appear in normal queries, resulting in false positives.
Malicious queries that contain rare or complex malicious code snippets do not match any rules, and might evade the detection, leading to false negatives.
This demonstrates the necessity of updating the WAF signature library, which can be achieved by acquiring the latest malicious queries via our system.

\emph{Code obfuscation.}
As mentioned above, we labeled unknown queries using static analysis, which might be bypassed by attacks using code obfuscation.
Code obfuscation could deteriorate the detection performance of static techniques, so we needed to eliminate the impact on AMODS.
To this end, we assumed that queries with obfuscated code were malicious and made no further analysis on them.
This operation is rational, because the reason for obfuscation in web requests is to evade detection, and it is futile to obfuscate benign web requests.
If necessary, dynamic analysis could be applied to defend against code obfuscation.
Dynamic techniques can monitor the code that is actually executed by the application, and thus are immune to many code-obfuscating transformations.

\emph{Web application concept drift.}
In machine learning, changes in the modeled behaviors of an observed object are known as concept drift \cite{Schlimmer1986Beyond}.
Maggi et al. \cite{Maggi2009Protecting} defined the behaviors of a web application as functionalities that the application offers and, more specifically, the content of web requests and responses.
They also defined the significant changes that frequently happen to behaviors of a web application as web application concept drift.
Web application concept drift is common nowadays because web applications become increasingly multi-featured to satisfy users' needs.
If web application concept drift is unknown to the web attack detection model, the model might misclassify the latest normal behaviors of the application after an update, leading to false positives.
Therefore, by keeping track of the FP rate of the daily-updated detection model on the original ten-day query set, our system could also be used to sense concept drift.
The FP rate increased dramatically on queries on the $6^{th}$ day; therefore it is likely that the web application concept drift occurred on this day.
We confirmed our speculation by analyzing logs and contacting the site manager.
The website was upgraded on the $6^{th}$ day: a new category for news was established, which caused the query structure in requests to the links in the new category to change, with new parameter names.

\emph{Future work.}
For future research, we intend to apply deep learning to discover more previously unknown attack features to replace $N$-gram features, since $N$-grams struggle to interpret attack patterns.
We also plan to enable AMODS to perform in a fully automatic manner without human supervision while maintaining its high detection rate, though at present it only requires a trivial amount of manual labeling work: 0.15\% (150 out of 100,000) of the entire unknown queries.
An intuitive solution is unsupervised learning.
Beyond that, AMODS is currently limited to detecting malicious queries in web attacks.
Nevertheless, it has excellent potential for use in network intrusion detection.
Comprehensive evaluation will be conducted to capture the performance of AMODS in this area.

\bibliographystyle{plain}

\bibliography{mybibfile}